\documentclass[preprint,12pt]{elsarticle}

\usepackage{amssymb}
\usepackage{amsmath}
\usepackage{subcaption}
\usepackage{booktabs}
\usepackage{color}
\usepackage{afterpage}
\journal{arXiv}

\begin{document}

\newcommand{\Li}[1]{\mathrm{Li}_{#1}}
\begin{frontmatter}



\title{Fast, Accurate Numerical Evaluation of Incomplete Planck Integrals}


\author{Whit Lewis and Ryan G.\ McClarren}

\affiliation{organization={Aerospace \& Mechanical Engineering, University of Notre Dame},%Department and Organization
            %addressline={Address One}, 
            city={Notre Dame},
            %postcode={00000}, 
            state={Indiana},
            country={USA}}

%\author[inst2]{Author Two}
%\author[inst1,inst2]{Author Three}

% \affiliation[inst2]{organization={Department Two},%Department and Organization
%             addressline={Address Two}, 
%             city={City Two},
%             postcode={22222}, 
%             state={State Two},
%             country={Country Two}}

\begin{abstract}
Methods for computing the integral of the Planck blackbody function over a finite spectral range, the so-called incomplete Planck integral, are necessary to perform multigroup radiative transfer calculations. We present a comparison, in terms of speed and accuracy, of a wide array of approaches to numerically evaluating these integrals. Our results indicate that a direct rational polynomial approximation to these integrals has the best combination of accuracy and efficiency. We also present for the first time a derivation of the polylogarithm form of these integrals and show that modern approaches to polylogarithm evaluation are suitable for numerically evaluating incomplete Planck integrals. This article is dedicated to Prof. B.D. Ganapol, the Transport Cowboy, on the occasion of his retirement.
\end{abstract}


%    \begin{highlights}
%    \item The first derivation of the polylogarithm form of the incomplete Planck integral is given.
%    \item A new rational approximation to these integrals is also given.
%    \item A computational study compares several different methods for evaluating these integrals in terms of speed and accuracy.
%    \end{highlights}


\end{frontmatter}


\section{Introduction}
\label{sec:Intro}
Radiative transfer calculations are important in a variety of physics applications. These include the modeling of inertial confinement fusion experiments \cite{Matzen:2005p2087,humbird2019transfer} and other high energy density phenomena \cite{Drake, Castor2004}. Benchmark solutions for several multifrequency radiative transfer problems have also been given by Ganapol (e.g., \cite{ganapolBlackbody}) and others \cite{mcclarren2021two}. 

In such high energy density radiative transfer calculations the linear transport equation for photons is coupled to a material energy equation through the absorption and emission of light. In local thermodynamic equilibrium, the emission of light is proportional to the Planck distribution, $B(\nu,T)$, a function of the medium temperature $T$ and the light frequency $\nu$. 
In these calculations often one must use a model of radiative transfer that accounts for the different photon energies. A popular approach to doing this is the multigroup method where one solves for integrals of the radiation intensity over energy.
In such a calculation, one must routinely compute integrals of the Planck function over a finite range of photon energies, also known as incomplete Planck integrals. These integrals are sometimes called ``band emission'' as well \cite{stewartBlackbody}.

The calculation of these integrals was treated in detail by Clark \cite{CLARK:1987il}. That work was first to the present, as a result from computer algebra software, that incomplete Planck integrals can be expressed as a combination of polylogarithm functions \cite{Lewin}. Clark then proceeds to use two approaches to calculate the incomplete Planck integral. The first uses a Taylor series representation of the integrand, and the other uses a power series representation of the polylogarithm. Each of these approaches has a limited range of validity and through an elegant analysis, Clark proposes a method that smoothly combines the two approaches to calculate Planck integrals.

In the intervening years there have been advances in calculating polylogarithm functions \cite{roughan2020polylogarithm,voigt2022comparison,voigt2023tri}. These have the potential to change the conclusions of Clark as to which calculation approaches are most efficient and accurate.

In this paper we compare recent approximations to polylogarithms, previously reported algorithms, and a new piecewise-rational polynomial approximation. Our analysis considers both the accuracy of the methods and the computational cost. We also give the first derivation of the polylogarithm form of the incomplete Planck integral whereas previous work \cite{CLARK:1987il, LampretBlackBody} relied on computer algebra software to realize this form {\em deus ex machina}.

 The paper is laid out as follows. In Section \ref{sec:planck} we detail the incomplete Planck integral that we are interested in calculating. We then give an overview of the theory of polylogarithms and their evaluation including their application to incomplete Planck integrals in Section \ref{sec:poly}. Section \ref{sec:results} compares the speed and accuracy of different approaches to evaluating Planck integrals, followed by a conclusions section. 

\section{The Planck Function and Its Integrals}
\label{sec:planck} 
\newcommand{\Enu}{E_\nu} 
The Planck function, $B(\nu,T)$, gives the intensity (or radiance) at a frequency $\nu$ of a blackbody at temperature $T$. This function is given by
\begin{equation}
    B(\nu,T) = \frac{2 h \nu ^3}{c^2} \left(e^{h\nu/kT} - 1 \right)^{-1},
\end{equation}where $h$ is the Planck constant, $k$ is the Boltzmann constant, and $c$ is the speed of light. The units of $B(\nu,T)$ are energy per area per steradian per time per frequency. Commonly, in high energy density systems, the Planck function is written in terms of photon energy, $\Enu = h\nu$. Because $B(\nu,T)$ is a distribution in frequency, we can convert it to a function of $\Enu$ via the relation
\begin{equation}
    B(\nu,T)\,d\nu = B(\Enu,T)\,d\Enu.
\end{equation}
Using $d\Enu / d\nu = h$, we get
\begin{equation}
    B(\Enu,T) = B(\nu,T)\,\frac{d\nu}{d\Enu} = \frac{2 \Enu^3}{h^3 c^2}\left(e^{\Enu/kT} - 1\right)^{-1}.
\end{equation}
Another way to write the Planck function is to define a reduced-frequency variable, $x = h\nu/kT$. This makes $dx/d\nu = h/kT$ and leads to
\begin{equation}
    B(x,T) = \frac{2 k^4 T^4}{h^3 c^2} \frac{x^3}{e^{x} - 1}.
\end{equation}

In any of the above forms the integral of the Planck function over all positive frequencies/energies surprisingly has a simple closed form. To see this we first note that 
\begin{equation}\label{eq:xpart}
    \int_0^\infty \frac{x^3}{e^{x} - 1} \, dx = \frac{\pi^4}{15}.
\end{equation}
This makes the integral of the Planck function 
\begin{equation}\label{eq:wholeInt}
    \int_0^\infty B(x,T)\, dx = \frac{\sigma_\mathrm{SB}}{\pi} T^4,
\end{equation}
where $\sigma_\mathrm{SB}$ is the Stefan-Boltzmann constant,
\begin{equation}
  \sigma_\mathrm{SB}=  \frac{2 \pi^5 k^4}{15 h^3 c^2} \approx 0.102831 \: \frac{\text{GJ}}{\text{cm}^2\cdot \text{ns}\cdot\text{keV}^4}.
\end{equation}
The result of this integral can also be written in terms of the radiation constant, 
\begin{equation}
    a_\mathrm{r} = \frac{4 \sigma_\mathrm{SB}}{c} \approx 0.01372 \: \frac{\text{GJ}}{\text{cm}^3\cdot\text{keV}^4},
\end{equation}
as 
\begin{equation}\label{eq:wholeInta}
    \int_0^\infty B(x,T)\, dx = \frac{a_\mathrm{r}c}{4\pi} T^4.
\end{equation}

The results in Eqs.~\eqref{eq:xpart} and \eqref{eq:wholeInt} naturally lead to the definition of a normalized Planck function 
\begin{equation}
    b(x) = \frac{15}{\pi^4} \frac{x^3}{e^{x} - 1},
\end{equation}
so that 
\begin{equation}
    B(x,T) = \frac{\sigma_\mathrm{SB}}{\pi}T^4 b(x).
\end{equation}
{
\color{black}
\subsection{Rosseland Function}
We also note that the derivative of $B(\nu,T)$ with respect to temperature is often needed for the linearization of the Planck spectrum and in diffusion approximations. This derivative, sometimes known as the Rosseland function, is written as
\begin{equation}
    R(\nu,T) \equiv\frac{\partial }{\partial T}B(\nu,T) = \frac{2 h^2 \nu ^4 e^{\frac{h \nu }{k T}}}{c^2 k T^2 \left(e^{\frac{h \nu }{k
   T}}-1\right)^2}.
\end{equation}
Changing variables to $x$ from $\nu$ using the same procedure as above leads to
\begin{equation}
    R(x,T) = \frac{4 \sigma_\mathrm{SB}}{\pi}T^3 r(x),
\end{equation}
where 
\begin{equation}
    r(x) = \frac{15}{4\pi^4} \frac{x^4 e^x}{(e^x-1)^2}.
\end{equation}
As we will see, integrals of $r(x)$ are related to integrals of $b(x)$.
}

\subsection{Incomplete Planck Integrals}
In many numerical methods for radiative transfer one needs to compute integrals of the Planck function over a finite range. This occurs in particular in the multigroup method where one  divides the frequency range into a finite number of intervals and integrates the radiation intensity over those intervals \cite{pomraning,mcclarren2021two}. That method requires computing integrals of the form
\begin{equation}\label{eq:bdefint}
   b_g\equiv \int_{x_g}^{x_{g+1}} b(x) \, dx = \Pi(b) - \Pi(a),
\end{equation}
where 
\begin{equation}
    \Pi(x) = \int_0^x b(z)\,dz.
\end{equation}
The values $x_g$ and $x_{g+1}$ are the reduced frequency values of the lower and upper bounds, respectively, of group $g$.

We should note that the function $\Pi(x)$ is tabulated in many heat transfer texts. Nevertheless, it does have a closed form expression first noted by Clark \cite{CLARK:1987il} and later by Lampret, et al. \cite{LampretBlackBody}. Both of those works used computer algebra software to perform the integration. Below we present a straightforward derivation of this formula. To find that expression will require becoming acquainted with a family of special functions called the polylogarithm.

\section{Polylogarithm Functions}\label{sec:poly}
Here we will sketch out the necessary theory of polylogarithms to compute the incomplete Planck integral, $\Pi(x)$. For a richer treatment of these functions, the reader should consult Lewin's monograph \cite{Lewin}.

One definition of the logarithm is given by the integral 
\begin{equation}
    -\log(1-z) = \int_0^z \frac{dt}{1-t}.
\end{equation}
This function also has a series representation given by
\begin{equation}\label{eq:log1mz}
     -\log(1-z) = \frac{z}{1} + \frac{z^2}{2} + \frac{z^3}{3} \dots, \qquad |z|\leq 1.
\end{equation}
The idea behind the dilogarithm, $\Li{2}(z)$, is to generalize these formulas. For the series representation we would make the power in the denominator 2 instead of 1:
\begin{equation}
    \Li{2}(z) = \frac{z}{1^2} + \frac{z^2}{2^2} + \frac{z^3}{3^2} + \dots, \qquad |z|\leq 1.
\end{equation}
Similarly, if we take Eq.~\eqref{eq:log1mz} and divide by z and integrate we get
\begin{align}
    -\int_0^z \frac{\log(1-t)}{t} &= \int_0^z dt\,\left(1 + \frac{t}{2} + \frac{t^2}{3} \dots\right) \nonumber \\
&= \left(\frac{z}{1^2} + \frac{z^2}{2^2} + \frac{z^3}{3^2} \dots\right) \nonumber \\
&= \Li{2}(z).
\end{align}
Note that while the series representation is only valid for $|z|\leq 1$, the integral can be defined for any value of $z$ by introducing branch cuts, etc.

This gives us a simple way to define a general polylogarithm $\Li{n}(z)$. We notice that the dilogarithm has the same series representation as $-\log(1-z)$ with the power in the denominator of each term increased from 1 to 2. Changing the power to another value, $n$, defines the polylogarithm
\begin{equation}
    \label{eq:poly}
\Li{n}(z) = \frac{z}{1^n} + \frac{z^2}{2^n} + \frac{z^3}{3^n} + \dots, \qquad |z|\leq 1.
\end{equation}
Note that $\Li{1}(z) = - \log(1-z)$. As already stated, the $\Li{2}(z)$ function is known as a dilogarithm. The third-order polylogarithm is known as the trilogarithm. Though the fourth-order version does not have a name, we dub it the tetralogarithm (pronouncing that word aloud perhaps explains why it has not been previously named). 

Using the same procedure as above (dividing by the argument and integrating) gives the integral definition of the polylogarithm:
\begin{equation}\label{eq:lin}
    \Li{n}(z) = \int_0^z \frac{\Li{n-1}(t)}{t}\,dt.
\end{equation}
From this definition we can show that $\Li{0}(z) = z/(1-z)$, and continue to determine polylogarithms of negative order. Additionally, we notice that the derivative of a polylogarithm can be determined from Eq.~\eqref{eq:lin} to be
\begin{equation}\label{eq:lderiv}
    \Li{n}'(z) = \frac{\Li{n-1}(z)}{z}.
\end{equation}

\begin{figure}[htbp]
    \centering
    % Subfigure 1
    \begin{subfigure}[b]{0.95\textwidth}
        \centering
        \includegraphics[width=\textwidth]{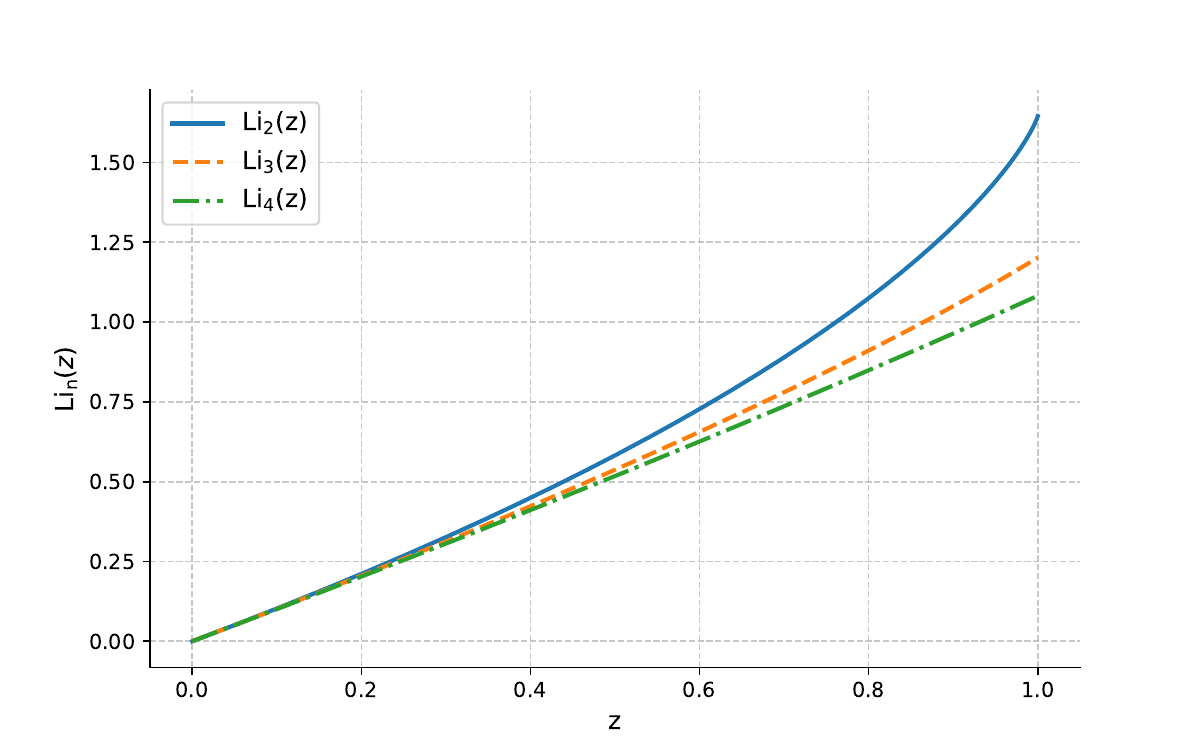} % Replace with your image
        \caption{$\Li{n}(z)$}
        \label{fig:Li}
    \end{subfigure}
    \hfill
    % Subfigure 2
    \begin{subfigure}[b]{0.95\textwidth}
        \centering
        \includegraphics[width=\textwidth]{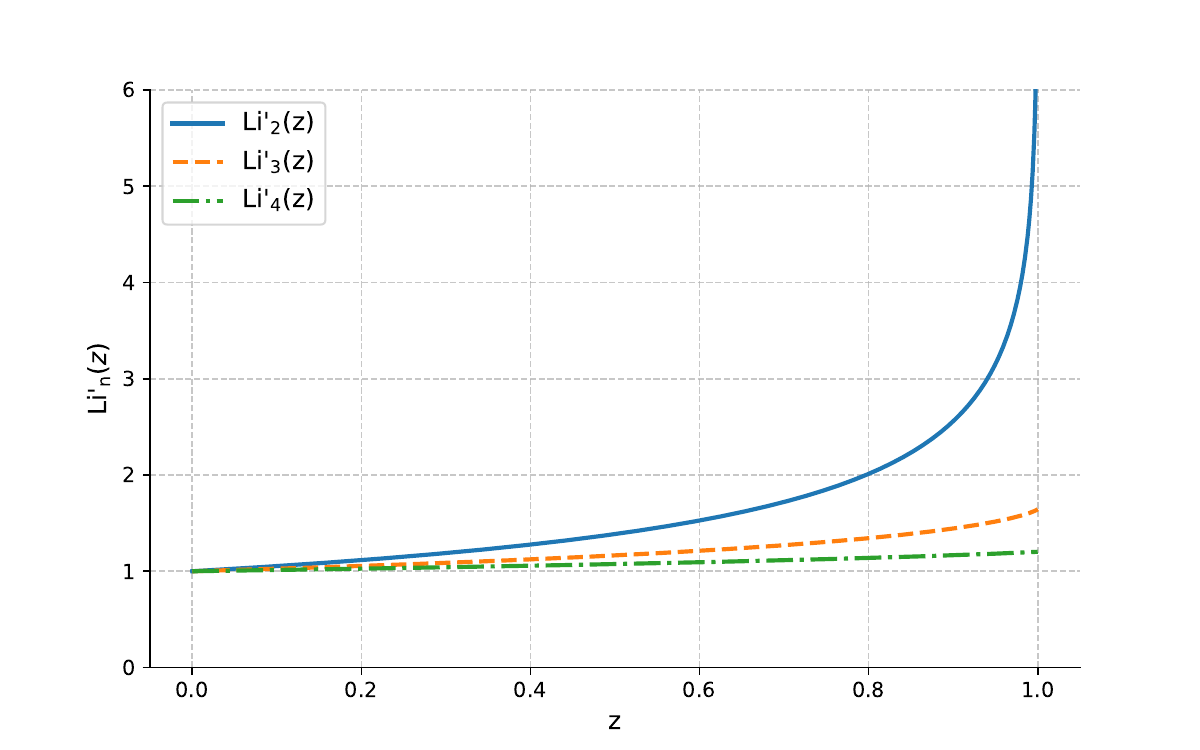} % Replace with your image
        \caption{$\Li{n}'(z)$}
        \label{fig:Liprime}
    \end{subfigure}
    % Main caption
    \caption{The first few polylogarithms and their derivatives over the range [0,1]. }
    \label{fig:polylogplot}
\end{figure}

The polylogarithm functions of orders 2 through 4 are plotted in Figure \ref{fig:polylogplot} along with their derivatives. In this figure we note that the higher-order polylogarithms are well-behaved functions in this range. The dilogarithm has its derivative go to infinity as $z \rightarrow 1$.

\subsection{Polylogarithms and Planck Integrals}
The detail of the polylogarithm functions above was necessary to understand how these functions are related to Planck integrals. We obtain the same result as \cite{CLARK:1987il}, but have considerably simpler algebra by resorting to complex numbers at the outset.

Taking the definition of $\Pi(x)$ from above and integrating once by parts we get
\begin{align}
    \Pi(x) &=\frac{15}{\pi^4} \int_{0}^x\frac{z^3}{e^{z} - 1} \,dz \nonumber \\ &=\frac{15}{\pi^4}\left(x^3\log \left(1-e^x\right)-x^4 - \int_0^x 3 z^2 \left(\log \left(1-e^z\right)-z\right)\,dz\right).
\end{align}
We can use the relation that $\log(1-e^x) = x + \log(1-e^{-x}) + \pi i$, where $i = \sqrt{-1}$, to get 
\begin{align}
    \Pi(x) &= \frac{15}{\pi^4}\left(x^3\left(\log(1-e^{-x}) + \pi i\right) - \int_0^x 3 z^2 \left(\log \left(1-e^{-z}\right)+ \pi i \right)\,dz\right) \nonumber \\ &= \frac{15}{\pi^4}\left(x^3\left(\log(1-e^{-x})\right)- \int_0^x 3 z^2 \log \left(1-e^{-z}\right)\,dz\right).  \label{eq:parts1} 
\end{align}
To continue we note that a simple application of the chain rule and Eq.~\eqref{eq:lderiv} leads to 
\begin{equation}
    \frac{d}{dz}\Li{n}(e^{-z}) = -\Li{n-1}(e^{-z}).
\end{equation}
Therefore, we can apply integration by parts again to the integral in Eq.~\eqref{eq:parts1} to get
\begin{equation}\label{eq:parts2}
    \int_0^x 3 z^2 \log \left(1-e^{-z}\right)\,dz = 3x^2\Li{2}(e^{-x}) - \int_0^x 6 z \Li{2}(e^{-z})\,dz.
\end{equation}
Finally, we apply integration by parts to the integral on the RHS of Eq.~\eqref{eq:parts2}
yielding
\begin{align}
    \int_0^x 6z \Li{2}(e^{-z})\,dz &= -6x\Li{3}(e^{-x}) + \int_0^x 6 \Li{3}(e^{-z})\,dz \nonumber \\
    &= -6x\Li{3}(e^{-x}) - 6\Li{4}(e^{-x}) + \frac{\pi^4}{15}.\label{eq:parts3}
\end{align}
This last integral used $\Li{4}(1) = \pi^4/90.$
Combining Eqs.~\eqref{eq:parts1}-\eqref{eq:parts3} gives
\begin{equation}
    \label{eq:pialmostfinal} \Pi(x) =  \frac{15}{\pi^4}\left( x^3\log(1-e^{-x}) - 3x^2\Li{2}(e^{-x})
    - 6x \Li{3}(e^{-x}) - 6 \Li{4}(e^{-x})\right) + 1.
\end{equation}
This is the equation that we need to numerically evaluate. Next, we discuss the subtleties of evaluating polylogarithms.
{\color{black}
\subsubsection{Integral of Rosseland function}
By a similar procedure we can integrate $r(z)$ from $z=0$ to $x$. The result is $\Pi(x)$ plus an additional term. Following \cite{CLARK:1987il}, we define the partial integral as 
\begin{align}
    \Upsilon(x) & \equiv \int_{0}^x r(z)\,dz \nonumber \\ &= 
    \Pi(x) - \frac{15}{4\pi^4} \frac{x^4}{e^x -1}.
\end{align}
Therefore, any procedure we have for evaluating $\Pi(x)$ can be used directly to evaluate $\Upsilon(x)$ because the additional term contains integer powers of $x$ and exponentials.
}

\subsection{Evaluating polylogarithms}
If we want to get numerical values for polylogarithms in the range $z\in [0,1]$, it seems natural to use the power series representation. However, as $z\rightarrow 1$ we would expect the series to converge slowly, especially for the dilogarithm due to the fact that its terms decay the slowest as the series becomes the harmonic series at $z=1$. This is important for evaluating Planck integrals because computing $\Pi(x)$ for $x$ near zero, we will be evaluating polylogarithms of the form $\Li{n}(e^{-x}) \rightarrow \Li{n}(1)$ as $x\rightarrow 0$.

Fortunately, there is an identity for the dilogarithm that can be used to evaluate near $z=1$. This relation \cite[Eq.~(1.11)]{Lewin} is
\begin{equation}\label{eq:li2trans}
    \Li{2}(z) = -\Li{2}(1-z) + \frac{\pi^2}{6} -  \log (1-z)\log z.
\end{equation}
The idea advanced in \cite{voigt2022comparison} is to find  a rational polynomial approximation to $\Li{2}(z)$ over the range $z\in[0,1/2]$ and use Eq.~\eqref{eq:li2trans} to evaluate $z\in(1/2,1]$. In \cite{voigt2022comparison} a degree $\{5,6\}$ rational polynomial is used. The approximation has a maximum relative error of about $5\times 10^{-17}$. The coefficients for the rational polynomial approximation are given in \ref{app:dilogarithm}

There are similar paths to evaluating the trilogarithm. We use the identity \cite[Eq.~(6.10)]{Lewin}
\begin{multline}
     \Li{3}(z) = -\Li{3}(1-1/z) - \Li{3}(1-z) + \\\zeta(3) + \textcolor{black}{\frac{\pi ^2}{6}}\log(z) - \frac{1}{2}\log^2(z)\log(1-z) + \frac{1}{6}\log^3(z), \label{eq:inversion2}
\end{multline}
where $\zeta$ is the Riemann zeta function; $\zeta(3) \approx 1.202056903$. This relation allows us to evaluate $\Li{3}(z)$ for $z\in(1/2,1]$ by evaluating a trilogarithm with a negative argument and a trilogarithm with argument in [0,1/2]. A rational polynomial approximation for each of these is given in \ref{app:trilogarithm}. The rational approximations are accurate to a similar level as that for the approximation to the dilogarithm.

Finally, we turn to the tetralogarithm. This function does not have as useful of a relation to shift the argument away from 1 as we have for the di- and trilogarithm. Nevertheless, the tetralogarithm function can be computed using 2 rational polynomial approximations \cite{voightCode} over the ranges $[0,1/2]$, $(1/2,8/10]$, and a 10-term power series for  $(8/10,1]$. These approximations are given in \ref{app:tetralogarithm}.

%==============================================
\section{Rational Approximations to the Incomplete Planck Integral} \label{sec:results}
%==============================================
As noted above, rational approximations to the individual polylogarithm functions have been successful in providing numerical values to double precision. However, to accomplish the calculation of $\Pi(x)$, inspection of Eq.~\eqref{eq:pialmostfinal} indicates we must evaluate at least one exponential, one logarithm, and three rational polynomials.  The reason for the caveat of ``at least'' is due to use of polylogarithm identities in Eqs.~\eqref{eq:li2trans} and \eqref{eq:inversion2} to expand the domain of the rational approximations.

An alternative is to construct a rational approximation for $\Pi(x)$ directly. Evaluating a rational polynomial would obviate the need to compute any exponentials or logarithms, potentially giving a large reduction in the evaluation time. Additionally, we can control the accuracy to match that of the direct evaluation of the polylogarithms discussed above. 

Inspired by rational approximations to the polylogarithms, we break the function $\Pi(x)$ up into 3 regions. The first region, from $x=0$ to 1, we use a polynomial to estimate the function. This polynomial has powers of $x$ from 3 to 12 and is not the Taylor series of $\Pi(x)$ to ensure near-uniform error over the range while maintaining the property of the function going to 0 at $x=0$.

\begin{figure}
    \centering
    \includegraphics[width=0.9\linewidth]{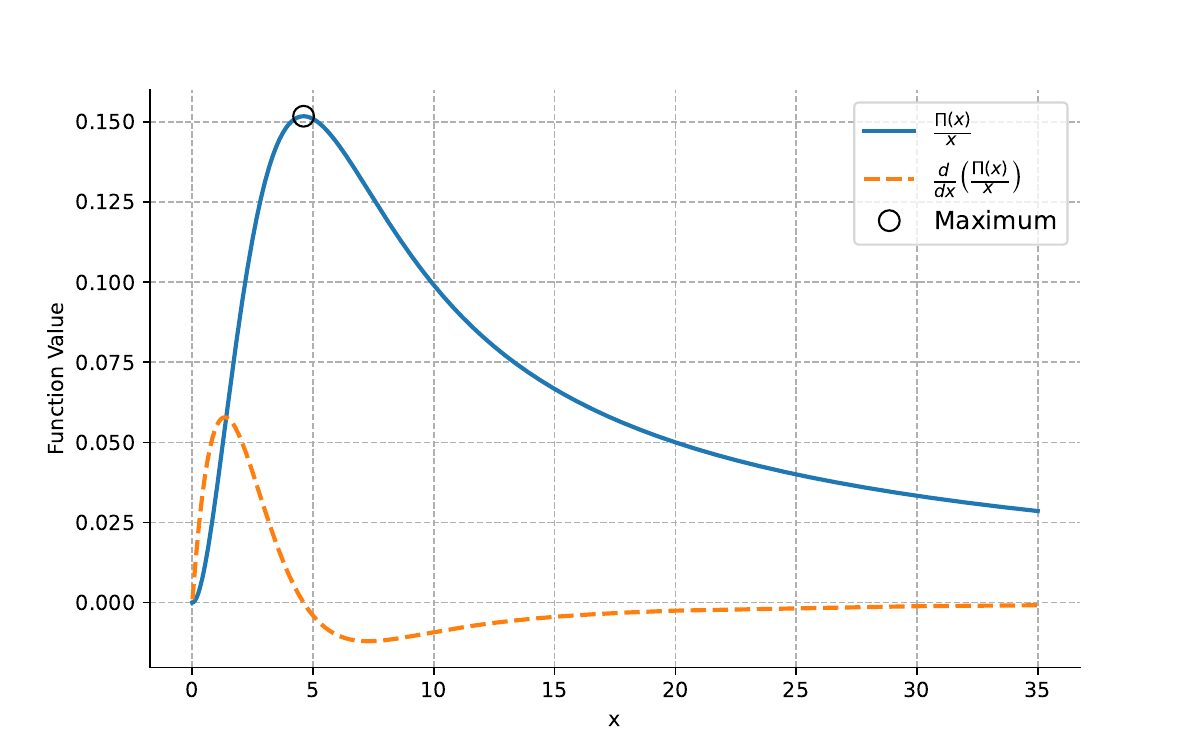}
    \caption{The function $\Pi(x)/x$ and its derivative. The maximum at $x_\mathrm{max}\approx 4.60834548$ is also indicated.}
    \label{fig:pix}
\end{figure}

We then break the domain of $x>1$ into two regions based on the behavior of $\Pi(x)/x$.  That function, as shown in Figure \ref{fig:pix}, will go to zero as $x\rightarrow \infty$ and that behavior can be engineered into the rational polynomial. The function $\Pi(x)/x$ also has a maximum, $x_\mathrm{max}$, at $x\approx 4.608345478$. This led us to break the domain above $x=1$ into two regions at this maximum.

The resulting approximation to $\Pi(x)$ is 
\begin{equation}\label{eq:piCases}
    \Pi_\mathrm{rp}(x) = \begin{cases}
        \displaystyle\sum_{n=3}^{12} \alpha_{n} x^n & 0\leq x\leq 1\\
        \displaystyle\frac{\displaystyle\sum_{n=0}^{14} p_{n} x^n}{\displaystyle\sum_{n=0}^{15} q_{n} x^n} & 1 < x\leq x_\mathrm{max}\\
        1+x\displaystyle\frac{\displaystyle\sum_{n=0}^{15} u_{n} x^n}{\displaystyle\sum_{n=0}^{18} v_{n} x^n} & x_\mathrm{max} < x
    \end{cases}.
\end{equation}
Note that when evaluating $\Pi(x_{g+1})-\Pi(x_g)$ as in Eq.~\eqref{eq:bdefint}, the $1$ for large $x$ cancels. In implementation, it is important to perform this cancellation algebraically to maintain precision as $\Pi(x)$ approaches 1. \textcolor{black}{Furthermore, we mention that in practice often the last group goes from $x = O(10)$ to $ x =\infty$. For this last group, we recommend that one use }
{\color{black}
\begin{equation}
    \Pi(x_G) - \Pi(x_{G-1}) = 1-\Pi(x_{G-1}),
\end{equation}
to take advantage of the known value of $\Pi(\infty)$.
}

\begin{table}[t]
  \centering
  \caption{Coefficients (\(\alpha_n\)) of the polynomial for the approximation to $\Pi(x)$ for small values of $x$.}
  \label{tab:polynomial_coefficients}
  \begin{tabular}{@{}l r@{}}
    \toprule
    Coefficient  & Value\\
    \midrule
    $\alpha_3$ & $ 5.13299112734207 \times 10^{-2}$ \\
    $\alpha_4$ & $ -1.92487167274314 \times 10^{-2}$ \\
    $\alpha_5$ & $ 2.56649556081953 \times 10^{-3}$ \\
    $\alpha_6$ & $ 3.38532276078157 \times 10^{-11}$ \\
    $\alpha_7$ & $ -3.05537299528005 \times 10^{-5}$ \\
    $\alpha_8$ & $ 7.73024978191323 \times 10^{-10}$ \\
    $\alpha_9$ & $ 5.64058261139052 \times 10^{-7}$ \\
    $\alpha_{10}$ & $ 2.48265186428919 \times 10^{-9}$ \\
    $\alpha_{11}$ & $ -1.37466236529825 \times 10^{-8}$ \\
    $\alpha_{12}$ & $ 1.08750502639625 \times 10^{-9}$ \\
    \bottomrule
  \end{tabular}
\end{table}
\begin{table}[t]
  \centering
  \caption{Coefficients of the numerator (\(p_n\)) and denominator (\(q_n\)) polynomials in Eq.~\eqref{eq:piCases} for approximating $\Pi(x)$ for $x \in (1,x_\mathrm{max}]$.}
  \label{tab:rational_polynomial_coefficients}
  \begin{tabular}{@{}l r@{}}
    \toprule
    Coefficient & Value \\
    \midrule
    $p_0$ & $-1.41249404058269 \times 10^{-19}$ \\
    $p_1$ & $ 2.10908974464015 \times 10^{-18}$ \\
    $p_2$ & $-1.50803009236045 \times 10^{-17}$ \\
    $p_3$ & $ 5.13299112734217 \times 10^{-2}$ \\
    $p_4$ & $ 2.43407740331316 \times 10^{-3}$ \\
    $p_5$ & $ 1.76278964709250 \times 10^{-3}$ \\
    $p_6$ & $ 1.84583709948367 \times 10^{-4}$ \\
    $p_7$ & $ 1.61066450038861 \times 10^{-6}$ \\
    $p_8$ & $ 5.55171714310303 \times 10^{-6}$ \\
    $p_9$ & $-4.01107999851710 \times 10^{-7}$ \\
    $p_{10}$ & $ 6.77941964950341 \times 10^{-8}$ \\
    $p_{11}$ & $-3.32386983347469 \times 10^{-9}$ \\
    $p_{12}$ & $ 2.16011320042569 \times 10^{-10}$ \\
    $p_{13}$ & $-4.41475932122409 \times 10^{-12}$ \\
    $p_{14}$ & $ 1.37839794361116 \times 10^{-13}$ \\
    \midrule
    $q_0$ & $ 1.00000000000000 \times 10^{0}$ \\
    $q_1$ & $ 4.22420253472632 \times 10^{-1}$ \\
    $q_2$ & $ 1.42749942493712 \times 10^{-1}$ \\
    $q_3$ & $ 3.60062420426830 \times 10^{-2}$ \\
    $q_4$ & $ 6.99146040951256 \times 10^{-3}$ \\
    $q_5$ & $ 1.18108372266407 \times 10^{-3}$ \\
    $q_6$ & $ 1.59466338512454 \times 10^{-4}$ \\
    $q_7$ & $ 1.88424240920010 \times 10^{-5}$ \\
    $q_8$ & $ 1.84136729899659 \times 10^{-6}$ \\
    $q_9$ & $ 1.53974339199716 \times 10^{-7}$ \\
    $q_{10}$ & $ 1.08080345539615 \times 10^{-8}$ \\
    $q_{11}$ & $ 6.36909463124141 \times 10^{-10}$ \\
    $q_{12}$ & $ 3.10142482314050 \times 10^{-11}$ \\
    $q_{13}$ & $ 1.21211579958412 \times 10^{-12}$ \\
    $q_{14}$ & $ 3.60098845901839 \times 10^{-14}$ \\
    $q_{15}$ & $ 8.40512459162060 \times 10^{-16}$ \\
    \bottomrule
  \end{tabular}
\end{table}

\begin{table}[t]
  \centering
  \caption{Coefficients of the numerator (\(u_n\)) and denominator (\(v_n\)) polynomials in Eq.~\eqref{eq:piCases} for approximating $\Pi(x)$ for $x > x_\mathrm{max}$.}
  \label{tab:rational_polynomial_uv}
  \begin{tabular}{@{}l r@{}}
    \toprule
    Coefficient & Value \\
    \midrule
    $u_0$ & $ 2.56572909391779 \times 10^{1}$ \\
    $u_1$ & $ -3.15998509049712 \times 10^{0}$ \\
    $u_2$ & $ -2.39927670638369 \times 10^{-1}$ \\
    $u_3$ & $ 8.58480323226633 \times 10^{-2}$ \\
    $u_4$ & $ -9.53770584997182 \times 10^{-3}$ \\
    $u_5$ & $ 6.40674177517660 \times 10^{-4}$ \\
    $u_6$ & $ -2.99266424885916 \times 10^{-5}$ \\
    $u_7$ & $ 1.02939262166793 \times 10^{-6}$ \\
    $u_8$ & $ -2.67840332691655 \times 10^{-8}$ \\
    $u_9$ & $ 5.32631745507925 \times 10^{-10}$ \\
    $u_{10}$ & $ -8.07975976849841 \times 10^{-12}$ \\
    $u_{11}$ & $ 9.21902324802904 \times 10^{-14}$ \\
    $u_{12}$ & $ -7.67993976921638 \times 10^{-16}$ \\
    $u_{13}$ & $ 4.41827357257837 \times 10^{-18}$ \\
    $u_{14}$ & $ -1.57179083187420 \times 10^{-20}$ \\
    $u_{15}$ & $ 2.60918226968872 \times 10^{-23}$ \\
    \midrule
    $v_0$ & $ 1.0 $ \\
    $v_1$ & $ -2.85898229933688 \times 10^{1}$ \\
    $v_2$ & $ 7.26162846427674 \times 10^{0}$ \\
    $v_3$ & $ -3.34829549552425 \times 10^{0}$ \\
    $v_4$ & $ 7.25820694406949 \times 10^{-1}$ \\
    $v_5$ & $ -1.62986793115613 \times 10^{-1}$ \\
    $v_6$ & $ 2.61495465830976 \times 10^{-2}$ \\
    $v_7$ & $ -3.72727464164911 \times 10^{-3}$ \\
    $v_8$ & $ 4.22901460469686 \times 10^{-4}$ \\
    $v_9$ & $ -4.08696607789196 \times 10^{-5}$ \\
    $v_{10}$ & $ 3.22894437288470 \times 10^{-6}$ \\
    $v_{11}$ & $ -2.13098148369549 \times 10^{-7}$ \\
    $v_{12}$ & $ 1.14549652872828 \times 10^{-8}$ \\
    $v_{13}$ & $ -5.01961354412149 \times 10^{-10}$ \\
    $v_{14}$ & $ 1.74053936269384 \times 10^{-11}$ \\
    $v_{15}$ & $ -4.67744488675722 \times 10^{-13}$ \\
    $v_{16}$ & $ 9.17011702895401 \times 10^{-15}$ \\
    $v_{17}$ & $ -1.20137760129307 \times 10^{-16}$ \\
    $v_{18}$ & $ 8.11756393563567 \times 10^{-19}$ \\
    \bottomrule
  \end{tabular}
\end{table}

 \afterpage{\clearpage}
%==============================================
\section{Other Approximations to the incomplete Planck integral}
%==============================================
\subsection{Combined Taylor/Power Series Approach}
The paper by Clark \cite{CLARK:1987il} contains several approximations to $\Pi(x)$ that can be numerically evaluated. The first of these uses a Taylor series approximation for $x$ having a small magnitude, and evaluates the power series definition of the polylogarithms given in Eq.~\eqref{eq:poly} for large values of $x$, and therefore small values of the argument $e^{-x}$ of the polylogarithms.

To switch between the approximations, Clark uses the following formula:
\begin{equation} \label{eq:clark_switch}
    \Pi_\mathrm{C}^{N,L}(x) = \mathrm{min}\left( \Pi_L(x), \tilde{\Pi}_{N}(x)\right).
\end{equation}
Here $\Pi_L(x)$ uses an $L$ term power series approximation for the polylogarithms (c.f., Eq.~\eqref{eq:poly}) in $\Pi(x)$, and $\tilde{\Pi}_{N}(x)$ is the $N$ term Taylor expansion of $\Pi(x)$ around $x=0$. This Taylor expansion is 
\begin{multline}
    \tilde{\Pi}_N(x) = \frac{15}{\pi^4}\left(\frac{x^3}{3} - \frac{x^4}{8} + \frac{x^5}{60} - \frac{x^7}{5040} + \frac{x^9}{272160} -\right.\\\left. \frac{x^{11}}{13305600} + \frac{x^{13}}{622702080} - \frac{691 x^{15}}
    {19615115520000} +\dots + c_N x^N\right).
\end{multline}
Note that in this approximation the number of terms is chosen to be $N =3, 5, 9, 13, 17,21,\dots$ so that the truncation error in the Taylor series is positive (i.e., $\tilde{\Pi}_N(x)$ goes to positive infinity as $x$ goes to infinity). This makes the selection of the minimum in Eq.~\eqref{eq:clark_switch} choose the approximation with the smaller error because $\Pi_L(z) < \Pi(z)$ for $z>1$.

\subsection{The ``Cold War'' approximations}
Zimmerman at Lawrence Livermore National Laboratory \cite{CLARK:1987il} and Goldin at the Keldysh Institute of Applied Mathematics of USSR Academy of Sciences \cite{chetverushkin1985mathematical} each developed fast approximations in the 1980s for the incomplete Planck integral. 

The Goldin approximation, $\Pi_\mathrm{G}$, is a simple, but effective approximation:
\begin{equation}
    \Pi_\mathrm{G}(x) = \begin{cases}
        \displaystyle \frac{15x^3}{\pi^4} \left(\frac{5 }{312}x^2-\frac{1}{8}x+\frac{1}{3}\right) & x\leq 2\\ \\
        \displaystyle 1 - \frac{15 }{\pi^4}e^{-x}\left(x^3+3 x^2+6 x+7.28\right) & x>2
    \end{cases}.
\end{equation}
The maximum relative error in this approximation is 0.274\% at $x\approx1.383$. 

The Zimmerman approximation, $\Pi_\frak{z}$, is more complicated, using more terms to achieve higher accuracy. Also, the branch taken depends on the upper bound of the range of the definite integral in Eq.~\eqref{eq:bdefint}, $b$.  This approximation is
\begin{equation}
\Pi_{\frak{z}}(x) =
\begin{cases} 
    \tilde{\Pi}_3(x) & \text{if } b < 10^{-3}, \\
    \displaystyle 1 - \frac{15}{\pi^4}\frac{ a_0 + a_1 x + a_2 x^2 + a_3 x^3 + a_4 x^4 + a_5 x^5}{1 + b_1 x + a_5 x^2} e^{-x} & \text{if } b \geq 10^{-3},
\end{cases}
\end{equation}
where
\begin{equation}
\begin{aligned}
    a_0 &= 6.493939402267, \\ a_1 &= 8.317008834543, \\
    a_2 &= 5.570970415031, \\ a_3 &= 2.161761553097, \\
    a_4 &= 0.5194172986679, \\ a_5 &= -0.07713864107538,\\
    b_1 &= 0.2807339758744.
\end{aligned} 
\end{equation}
The maximum relative error in this approximation is 0.07822\% at $x\approx 1.467$, about a factor of 3.5 better than $\Pi_\mathrm{G}$. 

\begin{figure}
    \centering
    \includegraphics[width=0.9\linewidth]{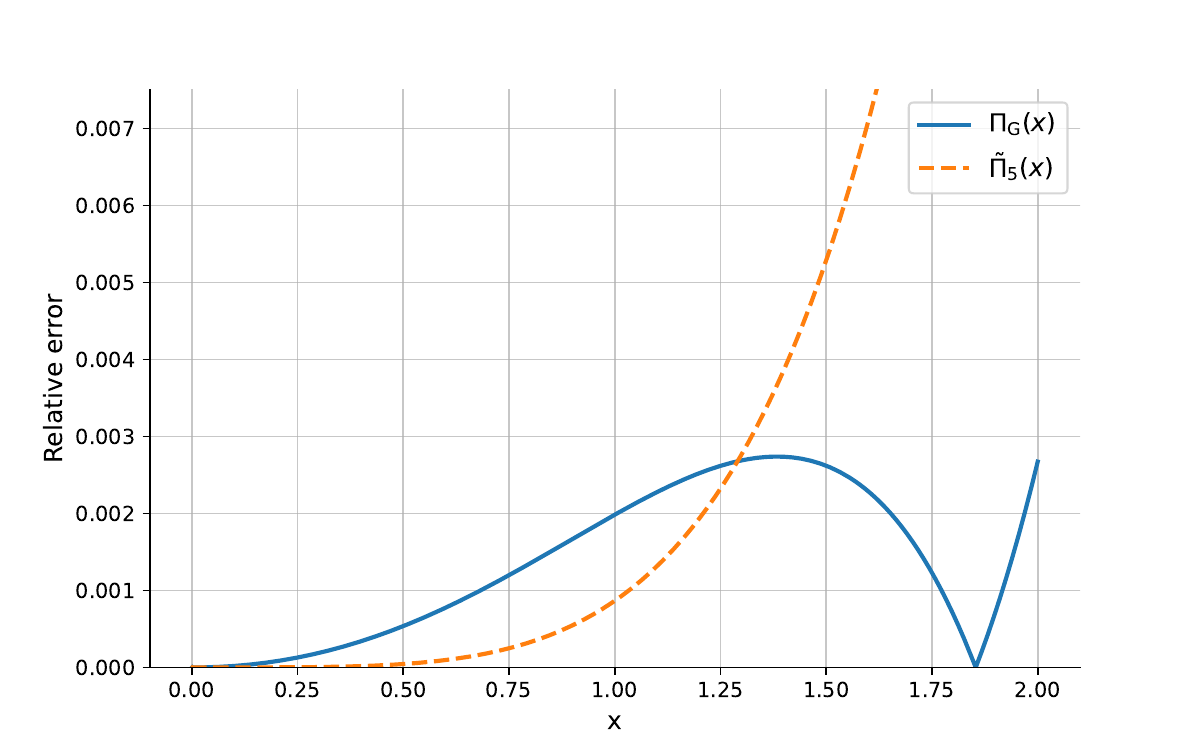}
    \caption{The relative error of the Goldin approximation for $x<2$ as compared to the Taylor series of degree 5, $\tilde{\Pi}_5$.}
    \label{fig:GoldinCompare}
\end{figure}

These two approximations both use a polynomial for low values, but the Goldin approach changes the coefficient for the $x^5$ term in the Taylor series from $1/60\approx 0.0166667$ to $5/312\approx 0.0160256$. This seemingly small change improves the behavior of the approximation for larger values of $x$, as shown in Figure \ref{fig:GoldinCompare}. 

\begin{figure}
    \centering
    \includegraphics[width=0.95\linewidth]{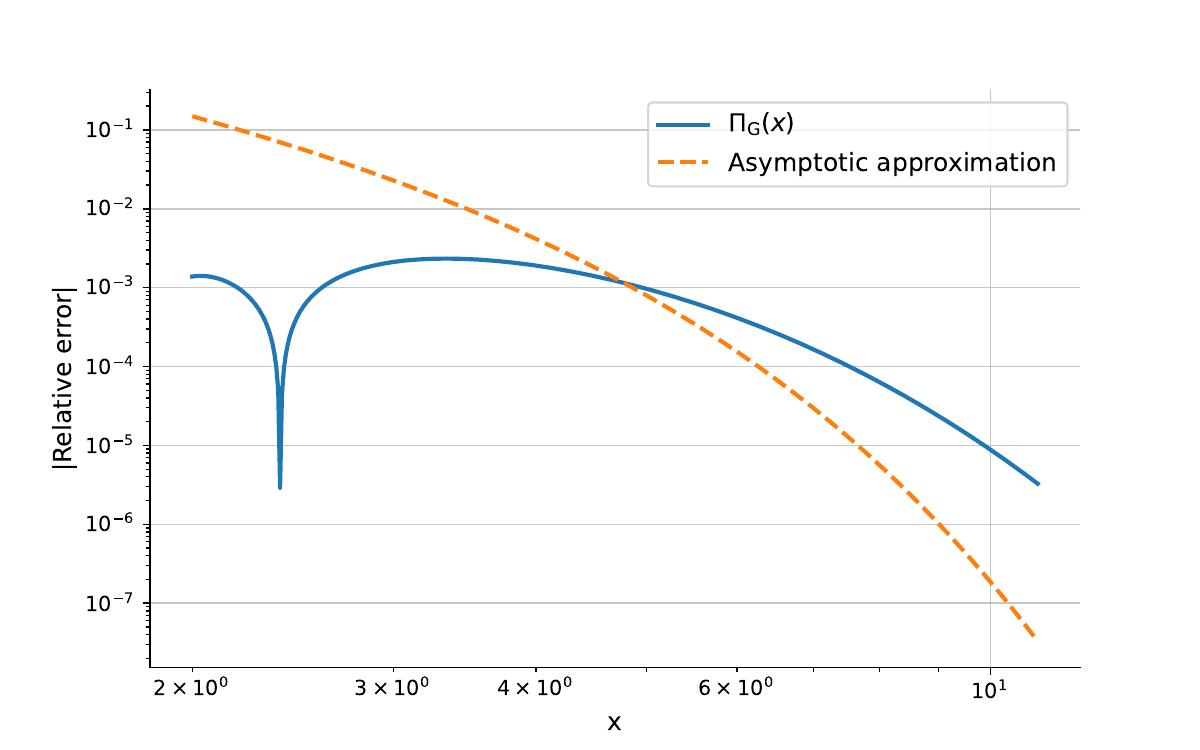}
    \caption{The relative error of the Goldin approximation for $x>2$ as compared to the asymptotic approximation.}
    \label{fig:GoldinCompareHigh}
\end{figure}

Additionally, the Goldin approximation for $x>2$ is based on the asymptotic approximation of $e^x-1 \approx e^x$ in the integrand and changing the bounds:
\begin{align}
    \Pi(x) &= \int_0^x \frac{15}{\pi^4} \frac{z^3}{e^z-1}\,dz \\ \nonumber
    &= 1 - \int^\infty_x \frac{15}{\pi^4} \frac{z^3}{e^z-1}\,dz \\ \nonumber
    &\approx 1 - \int^\infty_x \frac{15}{\pi^4} {z^3}e^{-z}\,dz\\ \nonumber
    &= 1 - \frac{15}{\pi^4} (x^3 + 3 x^2 + 6 x + 6) e^{-a}.
\end{align}
The Goldin approximation changes the constant in the polynomial multiplied by $e^{-x}$ from 6 to 7.28. This has the effect of making the approximation have a smaller error near $x=2$ by giving up a few digits of accuracy for $x$ large. This is shown in Figure \ref{fig:GoldinCompareHigh}.

\subsection{Quadrature-based Approximations}
Rather than approximating the function $\Pi(x)$ and then taking a difference to evaluate the incomplete Planck integral, a reasonable approach is to use quadrature rules directly on the integral of $b(x)$. Here we consider Gauss-Legendre quadrature rules shifted to the range of interest, i.e., 
\begin{equation}
    \label{eq:GLrule}
    \int_a^b b(x) \approx \sum_{n=1}^N w_n b(x_n),
\end{equation}
where the weights, $w_n$, and abscissae, $x_n$, are computed from the standard Gauss-Legendre quadrature sets with weights and abscissae $\{\hat{w}_n, z_n\}$ via
\begin{equation}
    \label{eq:weightsAb}
    x_n = \frac{1}{2}(z_n + 1)(b - a) + a, \qquad w_n = \frac{\hat{w}_n}{2} (b - a).
\end{equation}
We note that other types of quadrature would be possible, e.g., Romberg integration or tanh-sinh rules, but in our numerical experiments, Gauss-Legendre rules were superior in terms of accuracy per function evaluation.

When using quadrature rules, the accuracy of the calculation depends on the width of the range of integration. Therefore, the number of groups used in a radiative transfer calculation will determine the number of quadrature points used, and, therefore, the cost of computing the incomplete Planck integral. However, with more groups, as we will see, one can get a high level of accuracy with only a few quadrature points.

 \afterpage{\clearpage}
%==============================================
\section{Numerical Comparisons}
%==============================================
\subsection{Accuracy}
To test out the different approximations used to compute Planck integrals, we adapt the test problem used by Clark \cite{CLARK:1987il}. We first consider the question of accuracy. Here, we compute the incomplete Planck integral for a test problem for a medium with temperature $1$ keV and a variable number of groups, $G$, with the first group having the upper bound of $0.1$ keV and logarithmically spaced up to 20 keV and the last group extending from 20 keV to infinity. With group bounds denoted as $x_g$ for $g=0\dots G$, for any number of groups $x_0=0,$ $x_1=0.1$ keV, $x_{G-1} = 20$ keV, and $x_G = \infty$.  We compare the values in each group for $\Pi(x_{g+1}) - \Pi(x_g)$ to that calculated using arbitrary precision arithmetic in Mathematica. For the numerical calculations, to handle the highest energy group having a bound at infinity, we make the last group's value for the integral be evaluated so that the sum of all groups is 1.  For each method we compute the maximum relative error over all groups for each value of $G$. \textcolor{black}{We chose this approach to a test problem because it mimics what is commonly done when solving radiative transfer problems. An analyst would typically have a region of frequency space of interest and increase the number of groups to improve the resolution. Therefore, we want to give upper bounds on the accuracy of the different approaches in such a scenario.}

\begin{figure}
    \centering
    \includegraphics[width=0.95\linewidth]{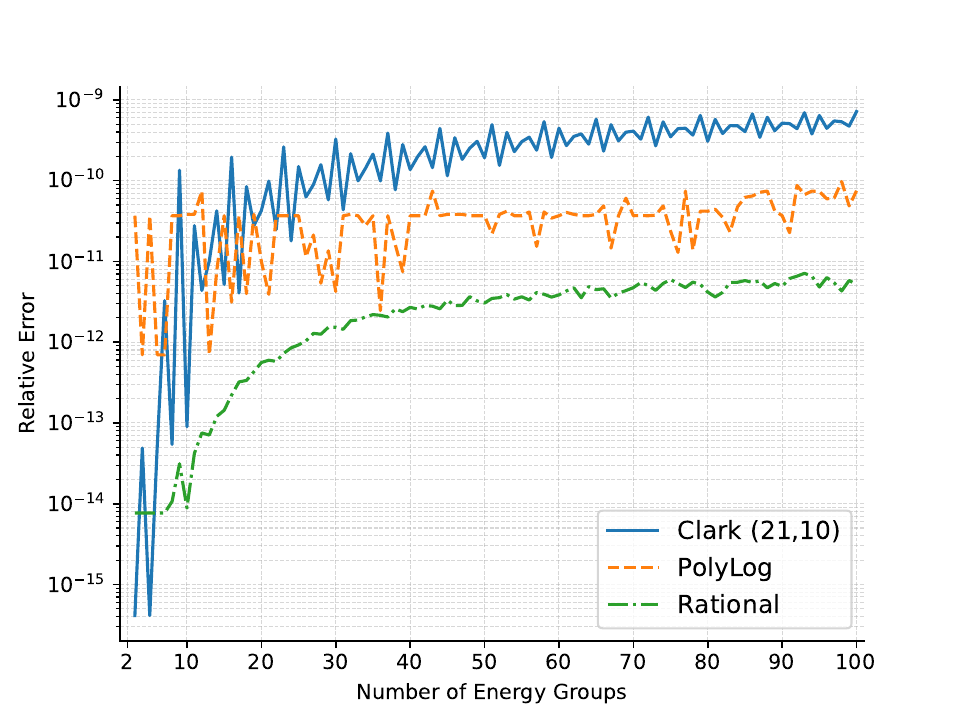}
    \caption{Comparison of the absolute value of the relative error for the three high accuracy methods on the test problem for different numbers of energy groups.}
    \label{fig:highacc}
\end{figure}

\begin{figure}
    \centering
    \includegraphics[width=0.95\linewidth]{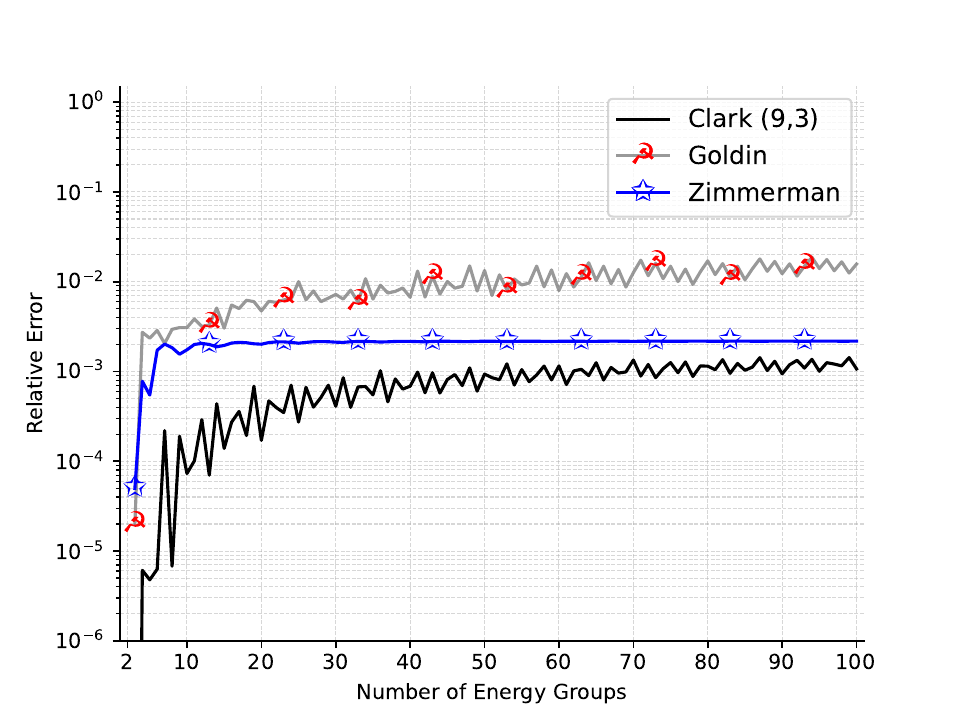}
    \caption{Comparison of the absolute value of the relative error for the three lower accuracy methods on the test problem for different numbers of energy groups.}
    \label{fig:lowacc}
\end{figure}

\begin{figure}
    \centering
    \includegraphics[width=0.95\linewidth]{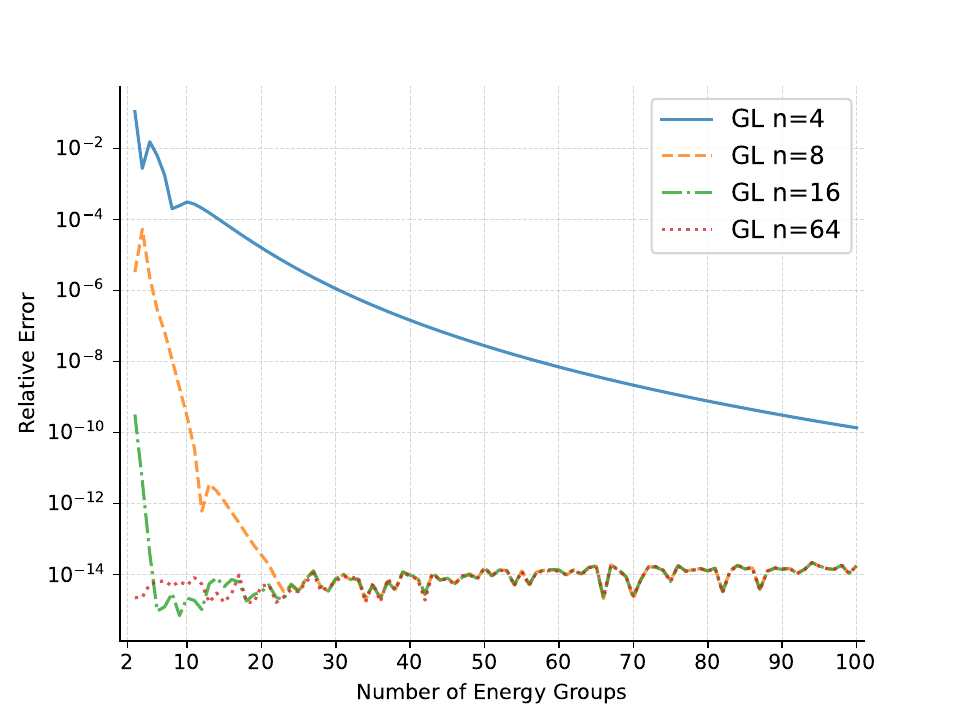}
    \caption{Comparison of the absolute value of the relative error for the Gauss-Legendre (GL) quadrature rules with varying number of quadrature points, $n$, on the test problem for different numbers of energy groups.}
    \label{fig:quadAcc}
\end{figure}

The different methods are compared in Figures \ref{fig:highacc}-\ref{fig:quadAcc}. In those figures, we denote the Clark approximation with the number of terms in the power series, $L$ and in the Taylor series, $N$, as ``Clark $N$-$L$''. The number of terms in the Clark approximation was chosen to match those recommended in \cite{CLARK:1987il}.

We first discuss the high-accuracy approximations to $\Pi(x)$ on our test problem. These approximations include the evaluation of the polylogarithms (``PolyLog'' in the figure), the rational approximation, and the Clark approximation $\Pi_\mathrm{C}^{21,10}$ recommended in \cite{CLARK:1987il}. From Figure \ref{fig:highacc} we can see that the high-accuracy approximations  have 10 digits of accuracy for a small number of groups. As the number of groups increases we observe that the rational approximation performs best, giving 11 digits of accuracy, followed by the polylogarithm evaluation with 10 digits, and finally $\Pi_\mathrm{C}^{21,10}$ with 9 digits. 

Next, we turn to the fast approximations of Goldin, Zimmerman, and $\Pi_\mathrm{C}^{9,3}$ in Figure \ref{fig:lowacc}. As anticipated by the discussion above, the Zimmerman approximation gives a lower error approximation than the Goldin formula; as the number of groups increases, the Zimmerman approximation gives 1 digit more accuracy over the Goldin formula. The few-term evaluation $\Pi_\mathrm{C}^{21,10}$ is about a factor of 2 better than the Zimmerman approximation in terms of relative error. 

Finally, the quadrature methods are compared in Figure \ref{fig:quadAcc}. Here we see the effect of group size on the error. For instance, below 10 groups, the 8 point quadrature rules have significant error that decreases by 5 orders of magnitude when the number of groups is 20 due to the shrinking of the energy group widths. Beyond 20 groups, 8 quadrature points is sufficient to guarantee accuracy up to double-precision. Nevertheless, with small numbers of energy groups more points (e.g., 64) may be required to get that level of accuracy. We also note that 4 quadrature points gives comparable accuracy to the fast approximations of $\Pi(x)$ and improves as the number of groups increases.


 \afterpage{\clearpage}
\subsection{Computational Speed}
To measure the speed of the calculations, we run a similar test problem in which we evaluate the incomplete Planck integral for $2\times 10^6$ different intervals. This is repeated 100 times to get a distribution of results. The different approximations are coded in Python using Numba \cite{numba} compilation\footnote{This code is freely available \cite{mccCode}.} using CPU threads to parallelize over the intervals. Additionally, Horner's method was used to efficiently evaluate all polynomials, and all quadrature sets were precomputed before begining the evaluations. The runs were completed on a AMD EPYC 7532 (2.4GHz/32-core) CPU in the University of Notre Dame Center for Research Computing.

\begin{figure}
    \centering
    \includegraphics[width=0.99\linewidth]{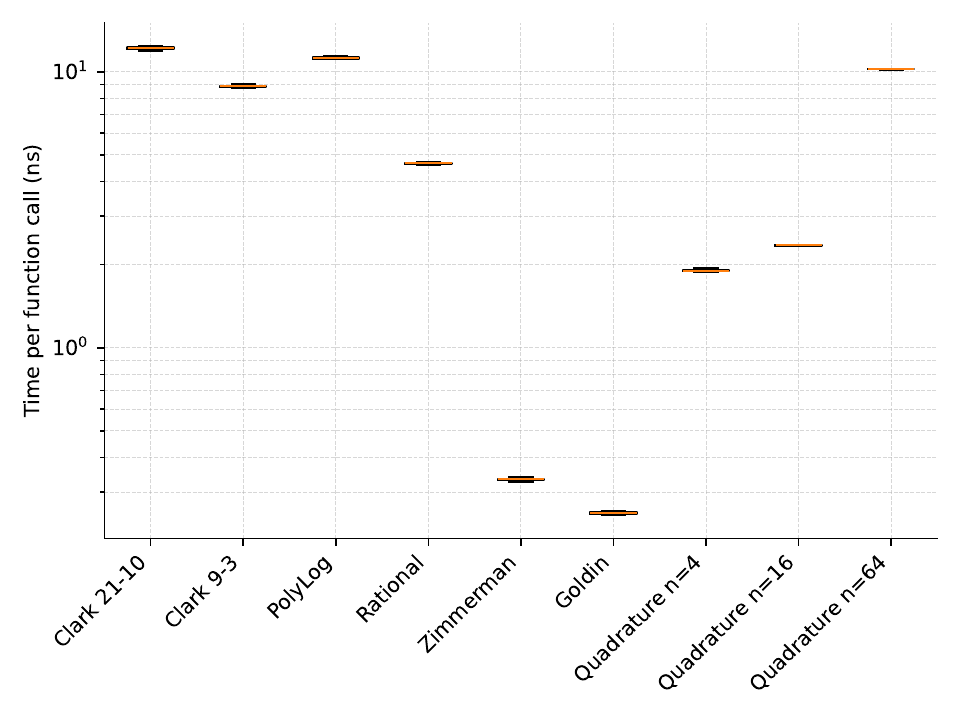}
    \caption{Box plot of the time per calculation of the incomplete Planck integral over 100 replicates of the test problem. Note that the inter-quartile range for each method is small enough to be indistinguishable from the mean.}
    \label{fig:timing}
\end{figure}

The means for time per function call over the 100 replicates is given in Table \ref{tab:method_times}. The table is sorted from longest running to fastest evaluation. Additionally, Figure \ref{fig:timing} is a boxplot of the results. The boxplot indicates that the variance in the distribution of times is small enough that all differences in the mean are statistically significant. From the table we can see that $\Pi_\mathrm{C}^{21,10}$ is the slowest evaluation. Directly evaluating the polylogarithms is faster. The next fastest was the 64-point quadrature  method. However, we find that the fastest among the high-accuracy methods is the rational approximation. This can, at least partially, be explained by the fact that it does not have to evaluate any exponentials or  logarithms, and, at worst, has to evaluate the quotient of a 15-term and an 18-term polynomial. 

From the table, we also observe that the fastest method is the Goldin approximation at about a quarter-nanosecond per evaluation. The next fastest is the Zimmerman approximation which, at 0.335 ns, is about 31\% slower than the Goldin evaluation. The slowest of the low-accuracy methods is the $\Pi_\mathrm{C}^{9,3}$ method which is 35x slower than Goldin. 
\begin{table}[t]
  \centering
  \caption{Timing data for the different methods in nanoseconds per function call.}
  \label{tab:method_times}
  \begin{tabular}{@{}l r@{}}
    \toprule
    Method & Time per function call (ns) \\
    \midrule
    Clark 21-10 $\Pi_\mathrm{C}^{21,10}$ & $12.196$ \\
    Polylogarithm  & $11.257$ \\
    Quadrature $n=64$ & $10.303$ \\
    Clark  9-3 $\Pi_\mathrm{C}^{9,3}$ & $8.915$ \\
    Rational & $4.648$ \\
    Quadrature $n=16$ & $2.438$ \\
    Quadrature $n=4$ & $2.014$ \\
    Zimmerman  & $0.335$ \\
    Goldin  & $0.255$ \\
    \bottomrule
  \end{tabular}
\end{table}

% Clark time 21-10: 1.219606399536133e-08
% Clark time 9-3: 8.91488790512085e-09
% PolyLog time: 1.1256814002990723e-08
% Rational time: 4.6480894088745115e-09
% Zimmerman time: 3.350973129272461e-10
% Goldin time: 2.54511833190918e-10
% Quadrature time n=4: 2.0144581794738766e-09
% Quadrature time n=16: 2.437770366668701e-09
% Quadrature time n=64: 1.0303080081939698e-08

\section{Conclusions}
We have presented a wide array of methods for evaluating incomplete Planck integrals. From our results on the accuracy and speed of the different methods we remark
\begin{itemize}
    \item The rational approximation we presented is the best combination of speed and accuracy of all the methods tested.
    \item Efficient, robust polylogarithm functions are available to directly evaluate the integral. This approach is faster and more accurate than the approximation $\Pi_\mathrm{C}^{21,10}$ that was recommended for high accuracy evaluation in \cite{CLARK:1987il}.
    \item The Goldin and Zimmerman methods are the two fastest evaluation methods with sub-nanosecond evaluation times. The Zimmerman method, while slower, does give an extra digit of accuracy.
    \item Gauss-Legendre quadrature can be an efficient and accurate approach, but the appropriate number of quadrature points depends on the width of the intervals.
\end{itemize}

We believe that the results above give clear guideposts for evaluating incomplete Planck integrals. If speed is of paramount importance, the Zimmerman and Goldin approaches are good choices. If high accuracy is desired, the rational approximations that we presented give a high degree of accuracy and speed. 

Further optimization of the calculations may be possible. For instance, we did not attempt to optimize the number of intervals in the rational approximations. For instance if, instead of 3 different approximations, we used 5 intervals, it is possible that the degree of the polynomials in the approximation could be reduced, decreasing computational cost. Furthermore, automatically choosing a number of  quadrature points based on the width of the group could prove to be the most efficient approach to high accuracy. This can be seen in our results because, for small enough group widths, the 16 point Gauss-Legendre quadrature gets 14 digits of accuracy and is nearly twice as fast as the rational approximation. However, to guarantee a particular accuracy, more points may be needed when the group bounds are wide.

{\color{black}
Finally, we mention that the Gauss-Legendre quadrature approach can also be beneficial when computing multigroup opacities with either the Planck or Rosseland weightings. For example, the Planck absorption opacity for group $g$ is
\begin{equation}
    \sigma_{\mathrm{a}g}^\mathrm{P}(T) = \frac{\displaystyle \int_{\nu_{g-1}}^{\nu_{g}} \sigma_\mathrm{a}(\nu,T) B(\nu,T)\,d\nu}{\displaystyle \int_{\nu_{g-1}}^{\nu_{g}}  B(\nu,T)\,d\nu}.
    \end{equation}
This function cannot be evaluated exactly, but the numerator can be evaluated using quadrature and the denominator using any of the methods discussed above. Similarly, for the Rosseland weighting,
\begin{equation}
    \sigma_{\mathrm{a}g}^\mathrm{R}(T) = \frac{\displaystyle \int_{\nu_{g-1}}^{\nu_{g}} \sigma_\mathrm{a}(\nu,T) \frac{\partial}{\partial T}B(\nu,T)\,d\nu}{\displaystyle \int_{\nu_{g-1}}^{\nu_{g}} \frac{\partial}{\partial T}B(\nu,T)\,d\nu},
    \end{equation}
the denominator is a difference of $\Upsilon(x_g) - \Upsilon(x_{g-1})$ and the numerator can be computed with Gauss-Legendre quadrature. Exploring the efficiency and accuracy of these approaches are natural follow-on work.
}
\section*{Acknowledgments}
This work was funded by the Center for Exascale Monte Carlo Neutron Transport (CEMeNT) a PSAAP-III project funded by the Department of Energy, grant number: DENA003967 and the National Science Foundation, grant number DMS-1906446.

\appendix

\section{Rational Approximations of Polylogarithm Functions}
The following gives approximations for the di-, tri-, and tetralogarithms based on the work of Voight \cite{voigt2022comparison, voigt2023tri}. Code in \texttt{C}, \texttt{C++}, and \texttt{Fortran} for these calculations is available \cite{voightCode}. Our Python translations of these methods is contained in \cite{mccCode}.

\subsection{Dilogarithm Calculations}
\label{app:dilogarithm}
Voight \cite{voigt2022comparison} uses the following rational approximation for the dilogarithm in the range $z\in[0,1/2]$:
\begin{equation}\label{eq:dirat}
    \Li{2}(z) \approx \frac{\displaystyle z \sum_{n=0}^5 p_n z^n}{\displaystyle \sum_{n=0}^6 q_n z^n}.
\end{equation}
The coefficients are given in Table \ref{tab:coeffs}. See Eq.~\eqref{eq:li2trans} to evaluate the dilogarithm in the range $z\in(1/2,1]$.

\begin{table}[t]
  \centering
  \caption{Coefficients of the numerator and denominator polynomials
    for the approximation of $\Li{2}(z)$ given in Eq.~\eqref{eq:dirat}.}
  \label{tab:coeffs}
  \begin{tabular}{@{}l r@{}}
    \toprule
    Coefficient & Value \\
    \midrule
    $p_0$ & $ 0.9999999999999999502 \times 10^{0}$ \\
    $p_1$ & $-2.6883926818565423430 \times 10^{0}$ \\
    $p_2$ & $ 2.6477222699473109692 \times 10^{0}$ \\
    $p_3$ & $-1.1538559607887416355 \times 10^{0}$ \\
    $p_4$ & $ 2.0886077795020607837 \times 10^{-1}$ \\
    $p_5$ & $-1.0859777134152463084 \times 10^{-2}$ \\
    \midrule
    $q_0$ & $ 1.0000000000000000000 \times 10^{0}$ \\
    $q_1$ & $-2.9383926818565635485 \times 10^{0}$ \\
    $q_2$ & $ 3.2712093293018635389 \times 10^{0}$ \\
    $q_3$ & $-1.7076702173954289421 \times 10^{0}$ \\
    $q_4$ & $ 4.1596017228400603836 \times 10^{-1}$ \\
    $q_5$ & $-3.9801343754084482956 \times 10^{-2}$ \\
    $q_6$ & $ 8.2743668974466659035 \times 10^{-4}$ \\
    \bottomrule
  \end{tabular}
\end{table}

\subsection{Trilogarithm Calculations}
\label{app:trilogarithm}
Voight \cite{voigt2023tri} uses two different rational approximations for the trilogarithm in the range $z\in[0,1/2]$ and $z\in[-1,0]$. Both have the form
\begin{equation}\label{eq:trirat}
    \Li{3}(z) \approx \frac{\displaystyle z \sum_{n=0}^5 p_n z^n}{\displaystyle \sum_{n=0}^6 q_n z^n}.
\end{equation}
The coefficients are given in Table \ref{tab:coeffs3neg} and \ref{tab:coeffs3half}. See Eq.~\eqref{eq:inversion2} to evaluate the trilogarithm in the range $z\in(1/2,1]$.

\begin{table}[t]
  \centering
  \caption{Coefficients of the numerator and denominator polynomials
    for the approximation of $\Li{3}(z)$ in the range $z\in[-1,0]$ given in Eq.~\eqref{eq:trirat}.}
  \label{tab:coeffs3neg}
  \begin{tabular}{@{}l r@{}}
    \toprule
    Coefficient & Value \\
    \midrule
    $p_0$ & $ 0.9999999999999999795 \times 10^{0}$ \\
    $p_1$ & $-2.0281801754117129576 \times 10^{0}$ \\
    $p_2$ & $ 1.4364029887561718540 \times 10^{0}$ \\
    $p_3$ & $-4.2240680435713030268 \times 10^{-1}$ \\
    $p_4$ & $ 4.7296746450884096877 \times 10^{-2}$ \\
    $p_5$ & $-1.3453536579918419568 \times 10^{-3}$ \\
    \midrule
    $q_0$ & $ 1.0000000000000000000 \times 10^{0}$ \\
    $q_1$ & $-2.1531801754117049035 \times 10^{0}$ \\
    $q_2$ & $ 1.6685134736461140517 \times 10^{0}$ \\
    $q_3$ & $-5.6684857464584544310 \times 10^{-1}$ \\
    $q_4$ & $ 8.1999463370623961084 \times 10^{-2}$ \\
    $q_5$ & $-4.0756048502924149389 \times 10^{-3}$ \\
    $q_6$ & $ 3.4316398489103212699 \times 10^{-5}$ \\
    \bottomrule
  \end{tabular}
\end{table}
\begin{table}[t]
  \centering
  \caption{Coefficients of the numerator and denominator polynomials
    for the approximation of $\Li{3}(z)$ in the range $z\in[0,1/2]$ given in Eq.~\eqref{eq:trirat}.}
  \label{tab:coeffs3half}
  \begin{tabular}{@{}l r@{}}
    \toprule
    Coefficient & Value \\
    \midrule
    $p_0$ & $ 0.9999999999999999893 \times 10^{0}$ \\
    $p_1$ & $-2.5224717303769789628 \times 10^{0}$ \\
    $p_2$ & $ 2.3204919140887894133 \times 10^{0}$ \\
    $p_3$ & $-9.3980973288965037869 \times 10^{-1}$ \\
    $p_4$ & $ 1.5728950200990509052 \times 10^{-1}$ \\
    $p_5$ & $-7.5485193983677071129 \times 10^{-3}$ \\
    \midrule
    $q_0$ & $ 1.0000000000000000000 \times 10^{0}$ \\
    $q_1$ & $-2.6474717303769836244 \times 10^{0}$ \\
    $q_2$ & $ 2.6143888433492184741 \times 10^{0}$ \\
    $q_3$ & $-1.1841788297857667038 \times 10^{0}$ \\
    $q_4$ & $ 2.4184938524793651120 \times 10^{-1}$ \\
    $q_5$ & $-1.8220900115898156346 \times 10^{-2}$ \\
    $q_6$ & $ 2.4927971540017376759 \times 10^{-4}$ \\
    \bottomrule
  \end{tabular}
\end{table}
\subsection{Tetralogarithm Calculations}
\label{app:tetralogarithm}
The tetralogarithm can be evaluated using either a rational approximation or a power series.
Voight \cite{voigt2023tri} uses two different rational approximations for the tetralogarithm in the range $z\in[0,1/2]$ and $z\in(1/2,8/10]$. These rational approximations are 
\begin{equation}\label{eq:tetra0}
    \Li{4}(z) \approx \frac{\displaystyle  z\sum_{n=0}^5 p_n z^n}{\displaystyle \sum_{n=0}^6 q_n z^n} \qquad z\in[0,1/2].
\end{equation}
\begin{equation}\label{eq:tetrahalf}
    \Li{4}(z) \approx \frac{\displaystyle  \sum_{n=0}^6 p_n z^n}{\displaystyle \sum_{n=0}^6 q_n z^n} \qquad z\in(1/2,8/10].
\end{equation}
The coefficients are given in Table \ref{tab:coeffs40} and \ref{tab:coeffs4half}.

The power series used for the range $(8/10,1]$ is given in \cite[Eq.~(4.2)]{roughan2020polylogarithm} as 
\begin{multline}
    \Li{4}(z) \approx \zeta(4) + \ell \zeta(3) + \frac{\ell^2}{2} \zeta(2) + \ell^3 \left( \frac{11}{36} - \frac{1}{6} \log(-\ell) \right) \\
-   \frac{\ell^4}{48} 
-\frac{\ell^5}{1440}
+  \frac{\ell^7}{604800} - \frac{\ell^9}{91445760}\quad z\in(8/10,1],
\end{multline}
where $\ell = \log z$.
\begin{table}[t]
  \centering
  \caption{Coefficients of the numerator and denominator polynomials
    for the approximation of $\Li{4}(z)$ in the range $z\in[0,1/2]$ given in Eq.~\eqref{eq:tetra0}.}
  \label{tab:coeffs40}
  \begin{tabular}{@{}l r@{}}
    \toprule
    Coefficient & Value \\
    \midrule
    $p_0$ & $ 1.0000000000000000414 \times 10^{0}$ \\
    $p_1$ & $-2.0588072418045364525 \times 10^{0}$ \\
    $p_2$ & $ 1.4713328756794826579 \times 10^{0}$ \\
    $p_3$ & $-4.2608608613069811474 \times 10^{-1}$ \\
    $p_4$ & $ 4.2975084278851543150 \times 10^{-2}$ \\
    $p_5$ & $-6.8314031819918920802 \times 10^{-4}$ \\
    \midrule
    $q_0$ & $ 1.0000000000000000000 \times 10^{0}$ \\
    $q_1$ & $-2.1213072418045207223 \times 10^{0}$ \\
    $q_2$ & $ 1.5915688992789175941 \times 10^{0}$ \\
    $q_3$ & $-5.0327641401677265813 \times 10^{-1}$ \\
    $q_4$ & $ 6.1467217495127095177 \times 10^{-2}$ \\
    $q_5$ & $-1.9061294280193280330 \times 10^{-3}$ \\
    \bottomrule
  \end{tabular}
\end{table}
\begin{table}[t]
  \centering
  \caption{Coefficients of the numerator and denominator polynomials
    for the approximation of $\Li{4}(z)$ in the range $z\in(1/2,8/10]$ given in Eq.~\eqref{eq:tetrahalf}.}
  \label{tab:coeffs4half}
  \begin{tabular}{@{}l r@{}}
    \toprule
    Coefficient & Value \\
    \midrule
    $p_0$ & $ 3.2009826406098890447 \times 10^{-9}$ \\
    $p_1$ & $ 9.9999994634837574160 \times 10^{-1}$ \\
    $p_2$ & $-2.9144851228299341318 \times 10^{0}$ \\
    $p_3$ & $ 3.1891031447462342009 \times 10^{0}$ \\
    $p_4$ & $-1.6009125158511117090 \times 10^{0}$ \\
    $p_5$ & $ 3.5397747039432351193 \times 10^{-1}$ \\
    $p_6$ & $-2.5230024124741454735 \times 10^{-2}$ \\
    \midrule
    $q_0$ & $ 1.0000000000000000000 \times 10^{0}$ \\
    $q_1$ & $-2.9769855248411488460 \times 10^{0}$ \\
    $q_2$ & $ 3.3628208295110572579 \times 10^{0}$ \\
    $q_3$ & $-1.7782471949702788393 \times 10^{0}$ \\
    $q_4$ & $ 4.3364007973198649921 \times 10^{-1}$ \\
    $q_5$ & $-3.9535592340362510549 \times 10^{-2}$ \\
    $q_6$ & $ 5.7373431535336755591 \times 10^{-4}$ \\
    \bottomrule
  \end{tabular}
\end{table}

 \bibliographystyle{elsarticle-num} 
 \bibliography{radtran}

\begin{thebibliography}{10}
\expandafter\ifx\csname url\endcsname\relax
  \def\url#1{\texttt{#1}}\fi
\expandafter\ifx\csname urlprefix\endcsname\relax\def\urlprefix{URL }\fi
\expandafter\ifx\csname href\endcsname\relax
  \def\href#1#2{#2} \def\path#1{#1}\fi

\bibitem{Matzen:2005p2087}
M.~Matzen, M.~Sweeney, R.~Adams, J.~Asay, Pulsed-power-driven high energy
  density physics and inertial confinement fusion research, Phys.~Plasmas
  12~(5) (2005) 055503.

\bibitem{humbird2019transfer}
K.~D. Humbird, J.~L. Peterson, B.~Spears, R.~G. McClarren, Transfer learning to
  model inertial confinement fusion experiments, IEEE Transactions on Plasma
  Science 48~(1) (2019) 61--70.

\bibitem{Drake}
R.~P. Drake, High Energy Density Physics, Springer-Verlag, New York, 2006.

\bibitem{Castor2004}
J.~Castor, Radiation hydrodynamics, Cambridge University Press, Cambridge,
  2004.

\bibitem{ganapolBlackbody}
B.~Ganapol, G.~Pomraning, {On the use of the Rosseland and Planck mean
  absorption coefficients in the non-equilibrium radiative transfer equation},
  Journal of Quantitative Spectroscopy and Radiative Transfer 37~(3) (1987)
  297--309.
\newblock \href {https://doi.org/10.1016/0022-4073(87)90050-1}
  {\path{doi:10.1016/0022-4073(87)90050-1}}.

\bibitem{mcclarren2021two}
R.~G. McClarren, Two-group radiative transfer benchmarks for the
  non-equilibrium diffusion model, Journal of Computational and Theoretical
  Transport 50~(6-7) (2021) 583--597.

\bibitem{stewartBlackbody}
S.~M. Stewart, {Blackbody radiation functions and polylogarithms}, Journal of
  Quantitative Spectroscopy and Radiative Transfer 113~(3) (2012) 232--238.
\newblock \href {https://doi.org/10.1016/j.jqsrt.2011.10.010}
  {\path{doi:10.1016/j.jqsrt.2011.10.010}}.

\bibitem{CLARK:1987il}
B.~A. Clark, {Computing Multigroup Radiation Integrals Using
  Polylogarithm-Based Methods}, Journal of Computational Physics 70~(2) (1987)
  311 -- 329.
\newblock \href {https://doi.org/10.1016/0021-9991(87)90185-9}
  {\path{doi:10.1016/0021-9991(87)90185-9}}.

\bibitem{Lewin}
L.~Lewin, Polylogarithms and associated functions, North Holland, 1981.

\bibitem{roughan2020polylogarithm}
M.~Roughan, The polylogarithm function in {Julia}, arXiv preprint
  arXiv:2010.09860 (2020).

\bibitem{voigt2022comparison}
A.~Voigt, Comparison of methods for the calculation of the real dilogarithm
  regarding instruction-level parallelism, arXiv preprint arXiv:2201.01678
  (2022).

\bibitem{voigt2023tri}
A.~Voigt, An algorithm to approximate the real trilogarithm for a real
  argument, arXiv preprint arXiv:2308.11619 (2023).

\bibitem{LampretBlackBody}
V.~Lampret, J.~Peternelj, A.~Krainer, {Luminous flux and luminous efficacy of
  black-body radiation: an analytical approximation}, Solar Energy 73~(5)
  (2002) 319--326.
\newblock \href {https://doi.org/10.1016/s0038-092x(02)00119-6}
  {\path{doi:10.1016/s0038-092x(02)00119-6}}.

\bibitem{pomraning}
G.~C. Pomraning, The Equations of Radiation Hydrodynamics, Dover Publications,
  Mineola, New York, 2005.

\bibitem{voightCode}
A.~Voight, \href{https://github.com/Expander/polylogarithm}{{Polylogarithm:
  Implementation of polylogarithms in C/C++/Fortran}}, accessed: 2024-12-05
  (2024).
\newline\urlprefix\url{https://github.com/Expander/polylogarithm}

\bibitem{chetverushkin1985mathematical}
B.~Chetverushkin, Mathematical modeling of problems in the dynamics of a
  radiating gas, Moscow Izdatel Nauka (1985).

\bibitem{numba}
H.~Finkel, S.~K. Lam, A.~Pitrou, S.~Seibert, {Numba}, Proceedings of the Second
  Workshop on the LLVM Compiler Infrastructure in HPC (2015).
\newblock \href {https://doi.org/10.1145/2833157.2833162}
  {\path{doi:10.1145/2833157.2833162}}.

\bibitem{mccCode}
R.~G. McClarren,
  \href{https://github.com/DrRyanMc/PlanckIntegrals}{{PlanckIntegrals: Python
  for evaluating incomplete Planck integrals}}, accessed: 2025-01-05 (2025).
\newline\urlprefix\url{https://github.com/DrRyanMc/PlanckIntegrals}

\end{thebibliography}





% \end{thebibliography}
\end{document}